\documentclass[aps,prl,preprintnumbers,twocolumn,groupedaddress,showpacs,superscriptaddress]{revtex4}
\usepackage{epsfig}
\usepackage{latexsym}
\usepackage{graphicx}
\usepackage{bm}
\usepackage{longtable}
\usepackage{color}

\newcommand{\kslash}{k\kern-1ex /}
\newcommand{\pslash}{p\kern-1ex /}
\newcommand{\qslash}{q\kern-1ex /}
\newcommand{\lslash}{l\kern-1ex /}
\newcommand{\sslash}{s\kern-1ex /}
\newcommand{\Dslash}{D\kern-1.2ex /}

\newcommand{\beqa}{\begin{eqnarray}}
\newcommand{\eeqa}{\end{eqnarray}}

\newcommand{\bd}{\begin{description}}
\newcommand{\ed}{\end{description}}

\newcommand{\ben}{\begin{eqnarray}}
\newcommand{\een}{\end{eqnarray}}

\def\lsim{\raise0.3ex\hbox{$<$\kern-0.75em\raise-1.1ex\hbox{$\sim$}}}
\def\gsim{\raise0.3ex\hbox{$>$\kern-0.75em\raise-1.1ex\hbox{$\sim$}}}
\def\simgt{\rlap{\lower 3.5 pt\hbox{$\mathchar \sim$}}\raise 1.0pt \hbox {$>$}}
\def\simlt{\rlap{\lower 3.5 pt\hbox{$\mathchar \sim$}}\raise 1.0pt \hbox {$<$}}

\begin{document}

\bibliographystyle{apsrev}

\preprint{UTCCS-P-57, UTHEP-601}

\title{Helium Nuclei in Quenched Lattice QCD}

\author{T.~Yamazaki}
\affiliation{Center for Computational Sciences, University of Tsukuba,
Tsukuba, Ibaraki 305-8577, Japan}
\author{Y.~Kuramashi}
\affiliation{Center for Computational Sciences, University of Tsukuba,
Tsukuba, Ibaraki 305-8577, Japan}
\affiliation{Graduate School of Pure and Applied Sciences,
University of Tsukuba, Tsukuba, Ibaraki 305-8571, Japan}
\author{A.~Ukawa}
\affiliation{Center for Computational Sciences, University of Tsukuba,
Tsukuba, Ibaraki 305-8577, Japan}
\collaboration{PACS-CS Collaboration}

\date{
%August       28, 2009
%October      30, 2009
%November      6, 2009
%November     16, 2009
%November     30, 2009, YK
%November     30, 2009
%December      5, 2009, AU
%December      6, 2009, YK
%December      7, 2009
\today
}

\begin{abstract}
We present results for the binding energies for He and $^3$He nuclei 
calculated in quenched lattice QCD at the lattice spacing of $a =0.128$ fm
with a heavy quark mass corresponding to $m_\pi = 0.8$ GeV.
Enormous computational cost for the nucleus correlation functions is
reduced by avoiding redundancy of equivalent contractions 
stemming from permutation symmetry of protons or neutrons in the 
nucleus and various other symmetries.
To distinguish a bound state from an attractive scattering state,
we investigate the volume dependence of the energy difference 
between the nucleus and the free multi-nucleon states by changing the
spatial extent of the lattice from 3.1 fm to 12.3 fm.
A finite energy difference left in the infinite spatial volume limit
leads to the conclusion that the measured ground states are bounded.
It is also encouraging that the measured binding energies and the experimental 
ones show the same order of magnitude. 
\end{abstract}

\maketitle

The atomic nuclei have been historically treated as collections of protons and 
neutrons.  The great success of the nuclear shell model since 
1949~\cite{{Mayer:1949xx},{Haxel:1949xx}}, 
explaining the nuclear magic numbers and detailed spectroscopy, has established 
that protons and neutrons are very good effective degrees of freedom 
at the nuclear energy scale of a few MeV.  
Nonetheless, 60 years later, we know for certain that protons and neutrons 
are made of quarks and gluons whose laws are governed by QCD.  
It is a great challenge to quantitatively understand the structure 
and property of known nuclei based on the first principle of QCD. 
This direct approach will be more important and indispensable if we are to 
extract reliable predictions for experimentally unknown nuclei in the 
neutron rich regions of the nuclear chart. 
In this article we address the fundamental question in the research in this 
direction, namely the binding energies of nuclei.

Interacting multi-baryon systems have been investigated by several 
studies in lattice QCD.
Nucleon-nucleon scattering was first studied in 
quenched QCD~\cite{{Fukugita:1994na},{Fukugita:1994ve}}. 
This work was followed by a partially-quenched mixed action 
simulation in Ref.~\cite{Beane:2006mx}.
Extraction of nuclear force between two nucleons has been investigated in
quenched and 2+1 flavor QCD~\cite{{Ishii:2006ec},{Aoki:2009ji},{Aoki:2008hh}}. 
All these studies assumed that the deuteron channel is not bound for 
the heavy pion mass, $m_\pi \simgt 0.3$ GeV, employed in the calculations.
Very recently, NPLQCD Collaboration has tried a feasibility study 
of the three-baryon system focusing on the quantum number 
of $\Xi^0\Xi^0 n$.  
They found the interaction to be repulsive~\cite{Beane:2009gs}.  
So far no evidence supporting bound state formation in multi-baryon systems 
has been observed in lattice QCD.
In this letter we examine the helium nuclei, He and $^3$He, in 
quenched lattice QCD using a heavy quark mass at a single lattice spacing.

The binding energy $\Delta E$ of the nucleus consisting 
of $N_N$ nucleons with the mass $m_N$ is very tiny compared 
with the mass $M$ of the nucleus: 
$\Delta E / M \sim O(10^{-3})$ with $\Delta E = N_N m_N-M$.
This causes a complicated situation 
that it is difficult to distinguish the physical binding energy
from the energy shift due to the finite volume effect 
in the attractive scattering system~\cite{Luscher:1986pf}.
One way to solve the problem is to investigate the volume dependence
of the measured  energy shift:
In the attractive scattering system 
the energy shift is proportional to $1/L^3$
at the leading order in the $1/L$ expansion~\cite{Luscher:1986pf,Beane:2007qr}, 
while the physical binding energy remains at a finite value 
at the infinite spatial volume limit.
In our simulation we choose three spatial extents corresponding to 
3.1, 6.1 and 12.3 fm, which are much larger than 
those employed in current numerical simulations so as to 
provide sufficient room for
the interacting multi-nucleon system.

A major computational problem with multi-nucleon systems in lattice QCD
is a factorially large number of Wick contractions of quark-antiquark fields 
required for evaluations of the nucleus correlation functions. 
A naive counting would give $(2N_p + N_n)!(2N_n + N_p)!$ for a nucleus composed of $N_p$ protons
and $N_n$ neutrons, which quickly becomes prohibitively large beyond 
three-nucleon system, {\it e.g., } 2880 for $^3$He and 518400 for He.

This number, however, contains equivalent contractions under the permutation 
symmetry in terms of the protons or the neutrons in the interpolating operator.
We can reduce the computational cost by avoiding the redundancy.
In case of the He nucleus which consists of the same number 
of the protons and the neutrons, the isospin symmetry 
also helps us reduce the necessary contractions.
After a scrutiny of the remaining equivalent contractions
by a computer we find that only 1107 (93) 
contractions are required for the He ($^3$He) nucleus correlation function. 
We have made a numerical test that the result with the reduced 
contractions reproduces the one with the full contractions on a
configuration.

Another technique to save the computational cost is 
a modular construction of the nucleus correlation functions.
We first make a block of three quark propagators where
a nucleon operator with zero spatial momentum is constructed 
in the sink time slice.
In this procedure we can incorporate the permutation symmetry
of two up (down) quarks in a proton (neutron) sink operator.
This is a simple trick to calculate $2^{N_N}$ contractions simultaneously. 
We also prepare several combinations of the two blocks 
which are useful for the construction of the nucleus correlators.

We carry out calculations on quenched configurations generated with the 
Iwasaki gauge action~\cite{Iwasaki:1983cj} at $\beta = 2.416$ whose
lattice spacing is $a=0.128$ fm determined with $r_0=0.49$ fm 
as an input~\cite{AliKhan:2001tx}.
We employ the HMC algorithm with the 
Omelyan-Mryglod-Folk integrator~\cite{{Omelyan:2003om},{Takaishi:2005tz}}.
The step size is
chosen to yield reasonable acceptance rate presented 
in Table~\ref{tab:conf_meas}.
We take three lattice sizes, 
$L^3\times T = 24^3 \times 64$, $48^3 \times 48$ and $96^3 \times 48$, 
to investigate the spatial volume dependence of the energy 
difference between the
nucleus and the free multi-nucleon states.
The physical spatial extents are 3.1, 6.1 and 12.3 fm, respectively.

\begin{table}[!t]
\caption{
Number of configurations ($N_{\rm conf}$), 
number of measurements on each configuration ($N_{\rm meas}$),
acceptance rate in the HMC algorithm,
pion mass ($m_\pi$) and nucleon mass ($m_N$).
\label{tab:conf_meas}
}
\begin{ruledtabular}
\begin{tabular}{cccccc}
$L$ & $N_{\mathrm{conf}}$ & $N_{\mathrm{meas}}$ 
& accept.(\%)
& $m_\pi$ [GeV] & $m_N$ [GeV]\\
\hline
24 & 2500 &  2 & 93 & 0.8000(3) & 1.619(2) \\
48 &  400 & 12 & 93 & 0.7999(4) & 1.617(2) \\
96 &  200 & 12 & 68 & 0.8002(3) & 1.617(2) \\
\end{tabular}
\end{ruledtabular}
\end{table}
We use the tadpole improved Wilson action 
with $c_{\mathrm{SW}} = 1.378$~\cite{AliKhan:2001tx}.
Since it becomes harder to obtain a reasonable signal-to-noise ratio at
lighter quark masses for the multi-nucleon system, 
we employ a heavy quark mass at $\kappa = 0.13482$ which gives
$m_\pi = 0.8$ GeV for the pion mass and $m_N = 1.6$ GeV for the nucleon mass. 
Statistics is increased by repeating the measurement of 
the nucleus correlation functions
with the source points in different time slices on each configuration.
The numbers for the configurations and the measurements on each configuration
are summarized in Table~\ref{tab:conf_meas}.
We separate 100 trajectories between each measurement 
with $\tau=1$ for the trajectory length.
The errors are estimated by the jackknife analysis choosing 200 
trajectories for the bin size.

The quark propagators are solved with
the periodic boundary condition 
in all the spatial and temporal directions,
and using the exponentially smeared source
$A\, e^{-B|{\vec x}|}$ after the Coulomb gauge fixing.
On each volume we employ two sets of the  
smearing parameters: $(A,B) = (0.5,0.5)$ and $(0.5,0.1)$ 
for $L=24$ and $(0.5,0.5)$ and $(1.0,0.4)$ for $L=48$ and 96.
Effective mass plots with different sources, which are shown later, 
help us confirm the ground state of the nucleus. 
Hereafter the first and the second smearing parameter 
sets are referred to as "$S_{1,2}$", respectively.

The interpolating operator for the proton is defined as 
$p_\alpha = \varepsilon_{abc}([u_a]^tC\gamma_5 d_b)u_c^\alpha$
where $C = \gamma_4 \gamma_2$ and $\alpha$ and $a,b,c$ are the Dirac index and
the color indices, respectively.
The neutron operator $n_\alpha$ is obtained 
by replacing $u_c^\alpha$ by $d_c^\alpha$ 
in the proton operator.
To save the computational cost
we use the nonrelativistic quark operator, in which the Dirac index
is restricted to upper two components.

The He nucleus has zero total angular momentum and positive parity
$J^P = 0^+$ with the isospin singlet $I = 0$. 
We employ the simplest He interpolating operator with the 
zero orbital angular momentum $L=0$, and hence $J=S$ with $S$ the total spin.
Such an operator was already given long time ago in Ref.~\cite{Beam:1967zz},
$\displaystyle{\mathrm{He} = 
\left( \overline{\chi}\eta - 
\chi \overline{\eta} \right) / \sqrt{2},
}$
where 
$
\chi$  =  ( $[+-+-]$ + $[-+-+]$ $- [+--+]$ $- [-++-]$ )/2 and 
$\overline{\chi}$  = ( 
$[+-+-]$ + $[-+-+]$ + $[+--+]$ + $[-++-]$$- 2 [++--]$$- 2 [--++]$ 
)/$\sqrt{12}$
with $+/-$ being up/down spin of each nucleon. 
$\eta, \overline{\eta}$ are obtained
by replacing $+/-$ in $\chi, \overline{\chi}$ by $p/n$ for the isospin.
Each nucleon in the sink operator is projected to have zero spatial
momentum.
We also calculate the correlation function of 
the $^3$He nucleus whose quantum numbers are 
$J^P=\frac{1}{2}^+$, $I = \frac{1}{2}$ and $I_z = \frac{1}{2}$.
We employ the interpolating operator in Ref.~\cite{Bolsterli:1964zz}
with the zero momentum projection on each nucleon in the sink operator.

\begin{figure}[!t]
\includegraphics*[angle=0,width=0.48\textwidth]{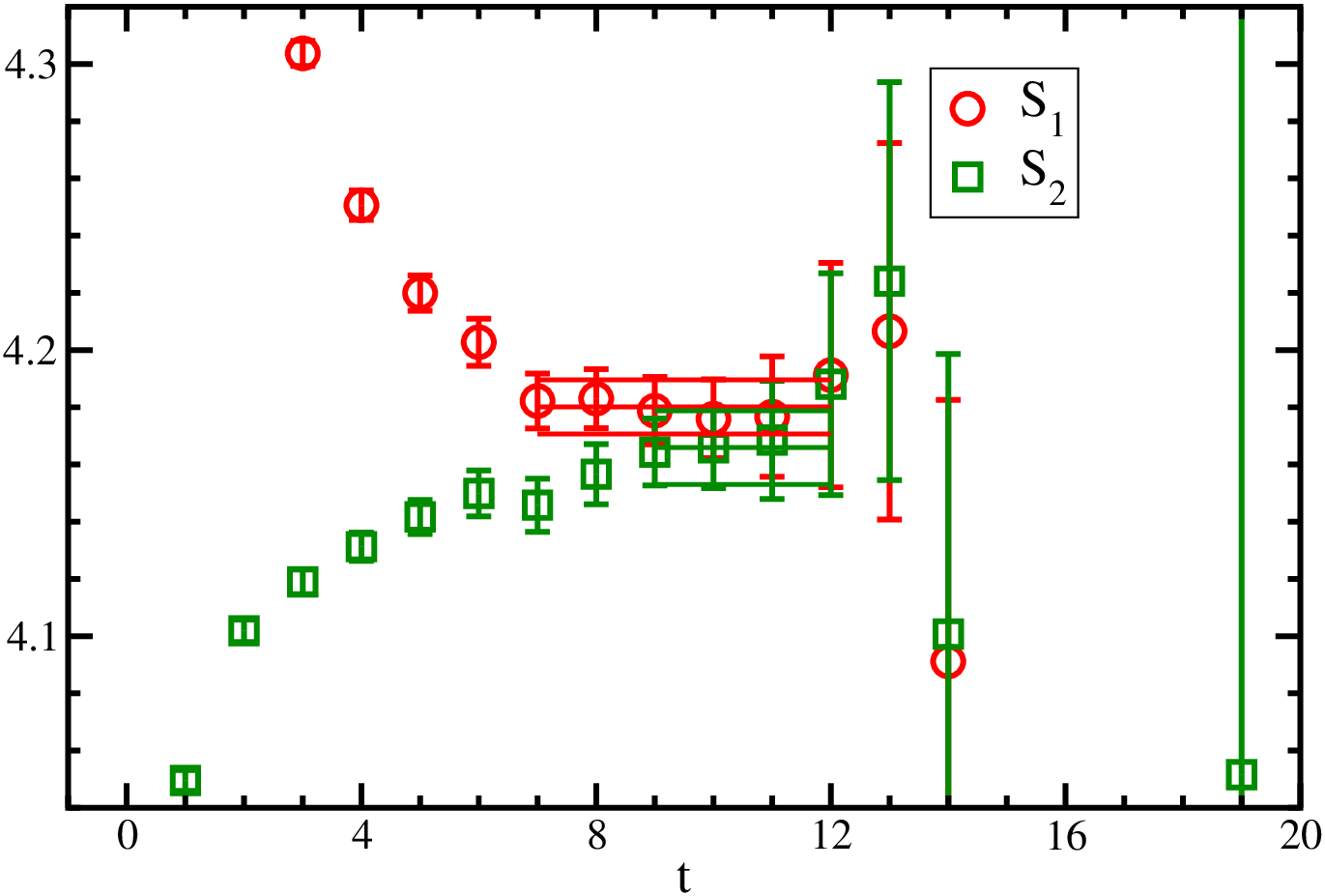}
\vspace{-0.3in}
\caption{
Effective mass of He nucleus with $S_1$ (circle) and $S_2$ (square) 
sources 
at $L=48$ in lattice units.
Fit results with one standard deviation error band are expressed 
by solid lines.
\label{fig:eff_He}
}
\end{figure}

Let us first present the He nucleus results.
Figure~\ref{fig:eff_He} shows the effective mass plots of 
the He nucleus correlators with the $S_{1,2}$ sources 
on the (6.1 fm)$^3$ spatial volume.
We find clear signals up to $t\approx 12$, beyond which
statistical fluctuation dominates.
The effective masses with the different sources show a reasonable
agreement in the plateau region. 
The consistency is also shown in the exponential fit results in
the plateau region as presented in the figure.

\begin{figure}[!t]
\includegraphics*[angle=0,width=0.48\textwidth]{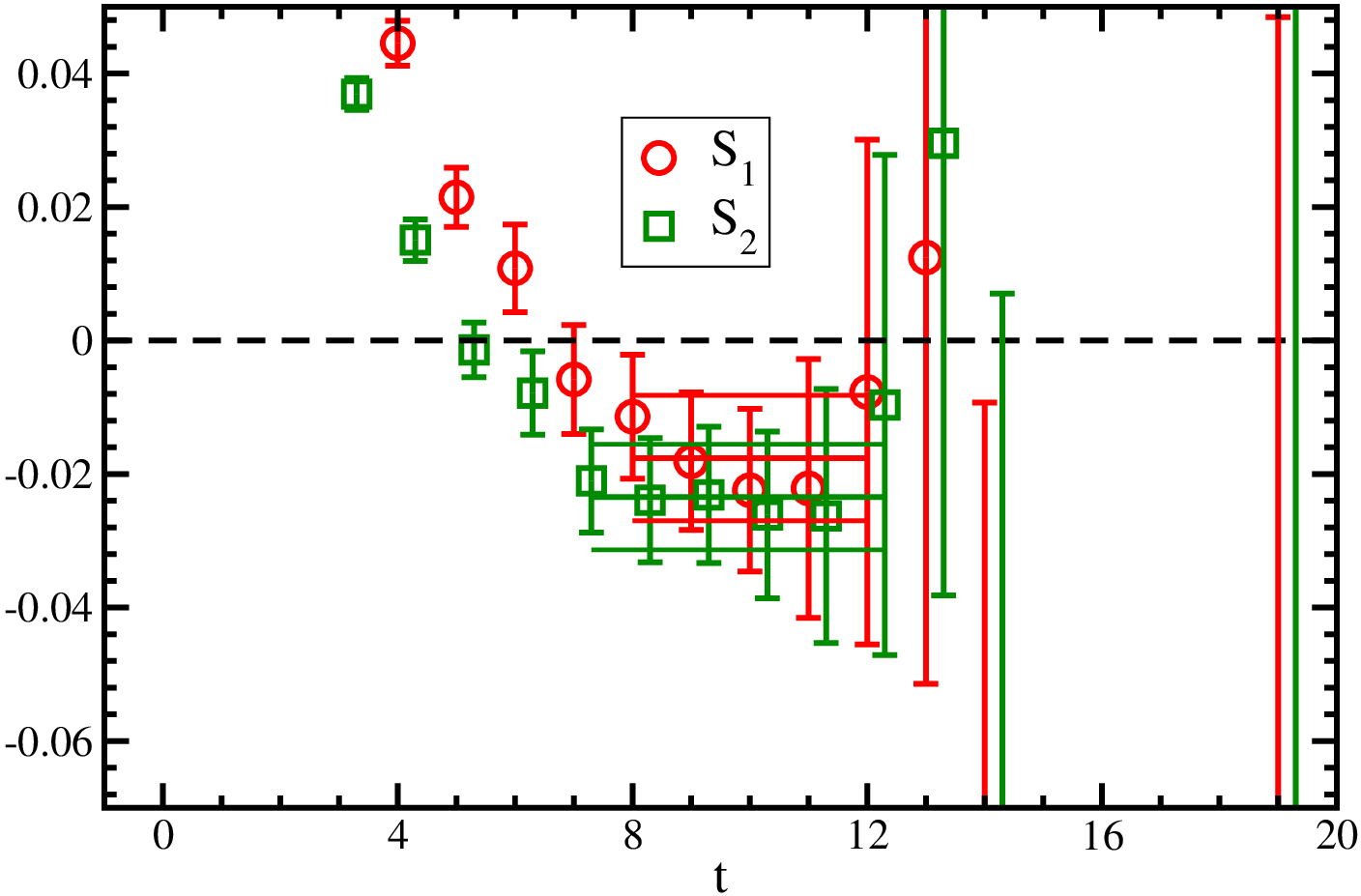}
\vspace{-0.3in}
\caption{
Effective energy shift of He nucleus in a convention of 
$-\Delta E_L^{\mathrm{eff}}$ 
with $S_1$ (circle) and $S_2$ (square) sources 
at $L=48$ in lattice units.
Square symbols are slightly shifted to positive direction in horizontal 
axis for clarity.
Fit results with one standard deviation error band are expressed 
by solid lines.
\label{fig:eff_R_He}
}
\end{figure}

In order to determine the energy shift $\Delta E_L$ precisely, 
we define the ratio of the He nucleus correlation function divided by
the fourth power of the nucleon correlation function,
$R(t) = G_{\mathrm{He}}(t)/(G_N(t))^4$, where $G_{\mathrm{He}}(t)$ 
and $G_N(t)$ are obtained with the same source.
The effective energy shift is extracted as
$\ln (R(t)/R(t+1)) = -\Delta E_L^{\mathrm{eff}}$,
once the ground states dominate in both correlators.
In Fig.~\ref{fig:eff_R_He} we present time dependence of 
$-\Delta E_L^{\mathrm{eff}}$ for the $S_{1,2}$ sources, both of which
show negative values 
beyond the error bars in the plateau region of $8 \le t \le 11$.
Note that this plateau region is reasonably consistent 
with that for the effective mass
of the He nucleus correlators in Fig.~\ref{fig:eff_He}.
The signals of $-\Delta E_L^{\mathrm{eff}}$ 
are lost beyond $t\approx 12$ because of 
the large fluctuations in the He nucleus correlators.
We determine $\Delta E_L$ by exponential fits of the ratios in 
the plateau region, $t=8-12$ for $S_1$ and 
$t=7-12$ for $S_2$, respectively.
We estimate a systematic error of $\Delta E_L$
from the difference of the central values of the fit results with
the minimum or maximum time slice changed by $\pm 1$.

\begin{figure}[!t]
\includegraphics*[angle=0,width=0.48\textwidth]{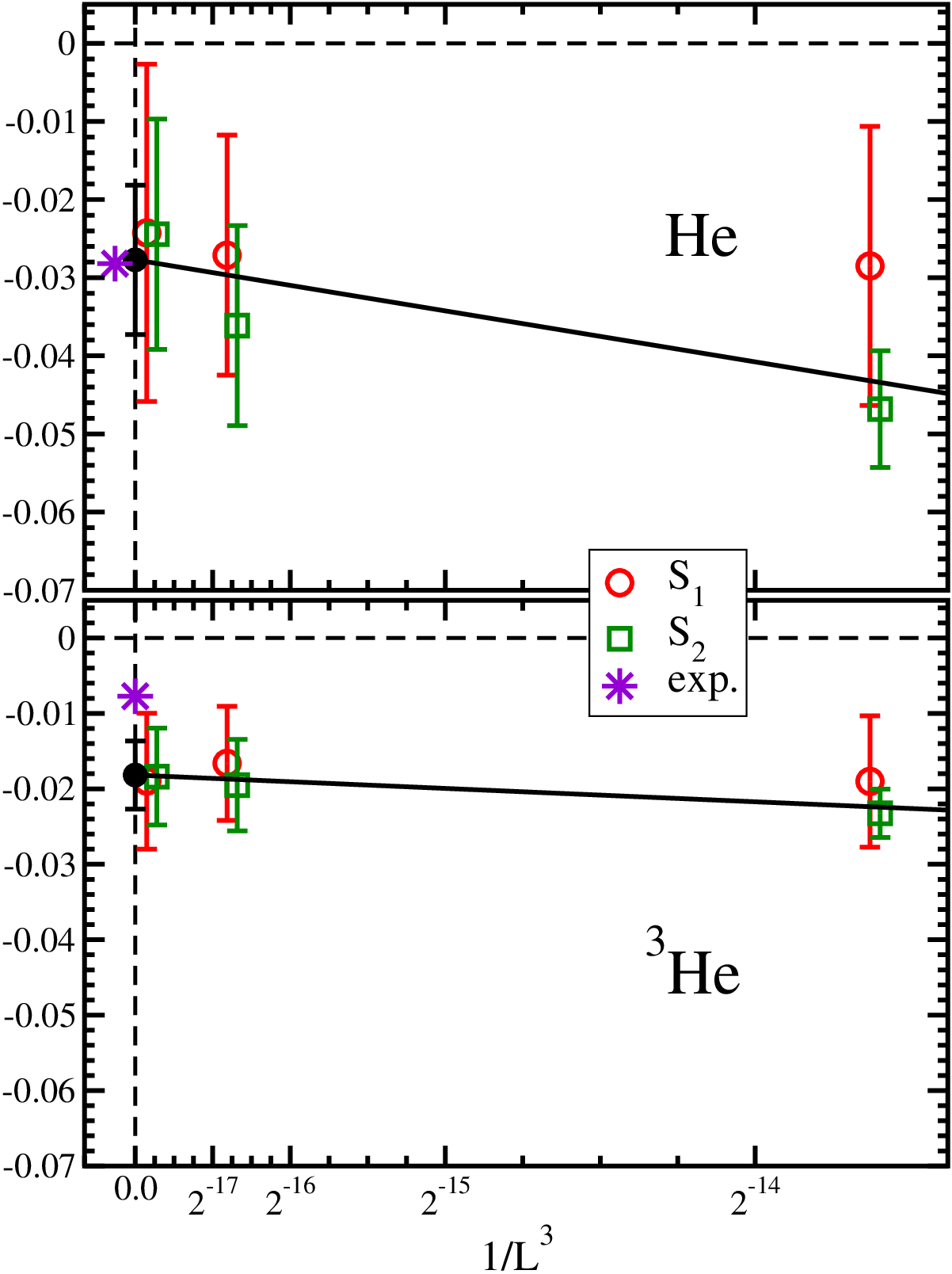}
\vspace{-0.3in}
\caption{
Spatial volume dependence of $-\Delta E_L = M - N_N m_N$ in GeV units
for He (upper) and $^3$He (lower) nuclei with $S_1$ (open circle) 
and $S_2$ (open square) sources.
Statistical and systematic errors are added in quadrature.
Square symbols are slightly shifted to positive direction in horizontal axis
for clarity. Extrapolated results to the infinite spatial 
volume limit (filled circle) 
and experimental values (star) are also presented.
\label{fig:dE_He}
}
\end{figure}

The volume dependence of the energy shift $\Delta E_L$
is plotted as a function of $1/L^3$ in the upper panel of 
Fig.~\ref{fig:dE_He}. Table~\ref{tab:dE} summarizes the numerical values
of $\Delta E_L$ at three spatial volumes, where the statistical
and systematic errors are presented in the first and second parentheses,
respectively.
The results for the $S_{1,2}$ sources are consistent within the error bars.
We observe little volume dependence for $\Delta E_L$ indicating 
a bound state, rather than the $1/L^3$ dependence expected for a 
scattering state, for the ground state in the He channel.

The physical binding energy $\Delta E$ defined 
in the infinite spatial volume limit
is extracted by a simultaneous fit of the data for the $S_{1,2}$ sources 
employing a fit function of $\Delta E + C/L^3$ with $\Delta E$ and $C$ 
free parameters.
The $1/L^3$ term is added to allow for contamination of scattering states.
A systematic error is estimated from the difference of the central values
of the fit results using the data with the different fit ranges
in the determination of $\Delta E_L$.
The result for $\Delta E$ is 0.0180(62)
in lattice units, which 
is 2.9 $\sigma$ away from zero as shown in Fig.~\ref{fig:dE_He}.
The error is evaluated from the statistical and systematic errors added 
in quadrature. In the following discussions we use the combined error.
We also try a pure bound state fit allowing for an exponentially small 
finite size correction:
$\Delta E$ and $\Delta E + C_1 e^{-C_2 L}$ with 
$\Delta E$ and $C_{1,2}$ free parameters.
We find all the results are in agreement with reasonable values of $\chi^2$.

Based on these analyses we conclude that the ground state of the measured 
four-nucleon system is bounded.  
An encouraging finding is that $\Delta E$ = 27.7(9.6) MeV 
with $a^{-1}=1.54$ GeV 
agrees with the experimental value of 28.3 MeV. 
However, we do not intend to stress the consistency because
our calculation is performed at the unphysically heavy 
pion mass, $m_\pi = 0.8$ GeV, and the electromagnetic interactions and
the isospin symmetry breaking effects are neglected.

\begin{table}[!t]
\caption{
\label{tab:dE}
Binding energies of He and $^3$He nuclei on each spatial volume. 
Extrapolated results to the infinite spatial volume limit 
are also presented. The first and second errors are
statistical and systematic, respectively.
}
\begin{ruledtabular}
\begin{tabular}{ccccc}
$L$ & \multicolumn{4}{c}{$\Delta E_L$ [MeV]}\\
 & He($S_1$) & He($S_2$) & $^3$He($S_1$) & $^3$He($S_2$) \\
\hline
24 & 28(14)(11) & 46.8(7.3)(1.6) & 19.0(6.3)(6.0) & 23.2(3.2)(0.5)\\
48 & 27(14)(05) & 36(12)(04)     & 16.6(6.9)(3.2) & 19.5(5.6)(2.3)\\
96 & 24(18)(12) & 24(14)(03)     & 19.0(7.6)(4.9) & 18.4(6.1)(1.9)\\
$\infty$ & \multicolumn{2}{c}{27.7(7.8)(5.5)} &
\multicolumn{2}{c}{18.2(3.5)(2.9)}\\
\end{tabular}
\end{ruledtabular}
\end{table}

We also calculate $\Delta E_L$ for the $^3$He nucleus with the $S_{1,2}$ 
sources,
whose results are presented in Fig.~\ref{fig:dE_He} and Table~\ref{tab:dE}. 
The trend of the volume dependence is similar to the He nucleus case.
A simultaneous fit of the data for the $S_{1,2}$ sources 
with a fit function of $\Delta E + C/L^3$
yields a finite value of $\Delta E= 18.2(4.5)$ MeV, 
which means the existence of a bound state
in the $^3$He nucleus channel. 
Our result for $\Delta E$ is about twice larger than 
the experimental value of 7.72 MeV. 
A main reason could be the heavy pion mass employed in this calculation.

As an alternative way to view this result, we compare the binding energies 
normalized by the atomic number: $\Delta E / N_N = 6.9(2.4)$ MeV and 
6.1(1.5) MeV for
the He and $^3$He nuclei, respectively.    
At our unphysically heavy pion mass, the three and four nucleon system 
does not show the experimental feature that the binding is stronger for 
He than for $^3$He.

We have addressed the issue of nuclear binding for the He and $^3$He 
nuclei.  We have shown that the current computational techniques and 
resources allow us to tackle this issue.  Albeit in quenched QCD and for 
unphysically heavy pion mass, we are able to extract evidence for 
the bound state nature of the ground state and the binding energies for 
these nuclei.

A future direction 
of primary importance is to investigate the quark mass dependence
of the binding energies of the nuclei.
There are several model studies of the quark mass dependence of the
nuclear binding energies~\cite{Flambaum:2007mj} which suggest that
the quark masses play an essential role in a quantitative understanding
of the binding energies. 
Another important issue is development of a strategy to calculate
nuclei with larger atomic numbers.
The required number of the Wick contractions quickly diverges
as the atomic number increases,
even if the redundancies are removed with various symmetries.
We leave it to future work.

Numerical calculations for the present work have been carried out
on the HA8000 cluster system at Information Technology Center
of the University of Tokyo and on the PACS-CS computer 
under the ``Interdisciplinary Computational Science Program'' of 
Center for Computational Sciences, University of Tsukuba. 
We thank our colleagues in the PACS-CS Collaboration for helpful
discussions and providing us the code used in this work.
This work is supported in part by Grants-in-Aid for Scientific Research
from the Ministry of Education, Culture, Sports, Science and Technology 
(Nos. 18104005, 18540250, 20105002, 21105501).
\bibliography{letter_he}

\end{document}